\documentstyle[]{article}
\begin{document}
\LARGE
\begin{center}
TWENTY-SEVEN LINES ON A CUBIC SURFACE AND HETEROTIC STRING SPACETIMES
\end{center}
\vspace*{0.3cm}
\Large
\begin{center}
Metod Saniga

\vspace*{.3cm}
\large
{\it Astronomical Institute, Slovak Academy of Sciences, 
SK-059 60 Tatransk\' a Lomnica,\\ The Slovak Republic}\\
(E-mail: msaniga@astro.sk -- URL: http://www.ta3.sk/$\widetilde{~}$msaniga)
\end{center}

\vspace*{-.4cm}
\noindent
\hrulefill

\vspace*{.2cm}
\noindent
\large
{\bf Abstract}\\
It is hypothetized that the algebra of the configuration of twenty-seven 
lines lying on a general cubic surface underlines the dimensional hierarchy
of heterotic string spacetimes.

\noindent
\hrulefill

\vspace*{.5cm}
\noindent
{\bf 1. Introduction}\\ 
The correct quantitative elucidation and deep qualitative understanding of
the observed dimensionality and signature of the Universe represent, 
undoubtedly, a crucial stepping stone on our path towards unlocking the
ultimate secrets of the very essence of our being. Although there have
been numerous attempts of a various degree of mathematical rigorosity and
a wide range of physical scrutiny to address this issue, the subject still
remains one of the toughest and most challenging problems faced by 
contemporary physics. In this contribution, we shall approach the problem by
raising a somewhat daring hypothesis that the dimensional aspect of the
structure of spacetime may well be reproduced by the algebra of a geometric
configuration as simple as that of the lines situated on a cubic surface in
a three-dimensional projective space.
\\ \\ 
{\bf 2. The Set of Twenty-Seven Lines on a Cubic Surface}\\
It is a well-known fact that on a generic cubic surface, $K_{3}$, there
is a configuration of twenty-seven lines /1/. Although this configuration
is geometrically perfectly symmetric as it stands, it exhibits a remarkable
non-trival structure when intersection/incidence relations between the
individual lines are taken into account. Namely, the lines are seen to form
three separate groups. The first two groups, each comprising six lines,
are known as {\it Schlafli's double-six}. This is indeed a remarkable subset
because the lines in either group are not incident with each other, i.e. they
are mutually skew, whereas a given line from one group is skew with one
and incident with the remaining five lines of the other group. The third
group consists of fifteen lines, each one being incident with four lines
of the Schlafli set and six other lines of the group in question. The basics
of the algebra can simply be expressed as:
\begin{equation}
27=12+15=2\times6+15.
\end{equation} 

There exists a particularly illustrative representation of this algebra. The
representation is furnished by a birational mapping between the points of 
$K_{3}$ and the points of a projective plane, $P_{2}$ /1/. Under such a 
mapping, the totality of the planar sections of $K_{3}$ has its 
counterpart in a linear, triply-infinite aggregate (the so-called web) of
cubic curves in $P_{2}$. Each cubic of the aggregate passes via six, 
generally distinct points B$_{i}$ ($i$=1,2,\ldots,6); the latter are called
the base points of the web. And the twenty-seven lines of $K_{3}$ are 
projected on $P_{2}$ as follows. The six lines $L_{i}^{(+)}$ (the first group
of Schlafli's double-six) are sent into (the neighbourhood of) the points
B$_{i}$. Other six lines $L_{j}^{(-)}$ ($j$=1,2,\ldots,6; the second 
Schlafli's 
group) answer to the six conics $Q_{j}$(B$_{1}$,B$_{2}$,\ldots,B$_{j-1}$,B$_{
j+1}$,\ldots,B$_{6}$), each passing via five of the base points. Finally, 
the remaining fifteen lines of the third group have their images in fifteen 
lines $L_{ij}$, joining the pairs of base points B$_{i}$B$_{j}$, $i \neq j$.
\\ \\ 
{\bf 3. An Algebra-Underlined Heterotic String Spacetime}\\
Now, let us hypothetise that the dimensional hierarchy of the Universe is
underlined by the above-discussed simple algebra, indentifying formally each 
line of $K_{3}$ with a single dimension of a heterotic string spacetime.
The total dimensionality of the latter would then be {\sf 27} instead of 26 
/2/. Further, we stipulate that the group of fifteen lines answers to
the first set of compactified dimensions of heterotic strings. We are thus
left with D$_{{\rm S}}$=12 dimensions corresponding to Schlafli's double-six,
and surmise that this ``Schlafli" spacetime is a natural setting for the 
M-theory, or, in fact, for the F-theory /3/; because our algebra also implies
that $12=2 \times 6=2 \times (5+1) = 10+2$!

And what about the four macroscopic dimensions familiar to our senses? A
hint for their elucidation may lie in the following observation. As 
explicitly pointed out, each line in the third group is incident with 
just {\it four} lines of the double-six. Let us assume that one of the
fifteen lines in this group has a special standing among the others; then
also the corresponding four Schlafli's lines have a distinguished footing
when compared with the rest in their group, and the same applies to the 
four dimensions they correspond to\ldots
 
To conclude, it is worth mentioning that our hypothesis gets a significant
support from a recent finding by El Naschie /4/, based on the so-called
Cantorian fractal-space approach, that the exact Hausdorff dimension of
heterotic string spacetimes is 26.18033989, i.e. {\it greater than} 26.
\\ \\ 
{\bf References}

\normalsize
\vspace*{.1cm}
\noindent
1. A. Henderson: 1911. {\it The 27 lines upon a cubic surface}, Cambridge
University Press, Cambridge.

\vspace*{.1cm}
\noindent
2. M. B. Green, J. H. Schwartz and E. Witten: 1987, {\it Superstring theory},
Cambridge University Press, 

Cambridge.

\vspace*{.1cm}
\noindent
3. M. Kaku: 1999, {\it Introduction to superstrings and M-theory}, Springer
Verlag, New York.

\vspace*{.1cm}
\noindent
4. M. S. El Naschie: 2001, ``The Hausdorff dimensions of heterotic string 
fields are D$^{(-)}$=26.18033989 

and D$^{(+)}$=10," {\it Chaos, Solitons $\&$
Fractals}, {\bf 12}, 377--379.

\end{document}